# What do neuroanatomical networks reveal about the ontology of human cognitive abilities?

Daniel Kristanto[1], Xinyang Liu[2,3], Werner Sommer[4,5], Andrea Hildebrandt[2,6*#], Changsong Zhou[1,7*#]

1: Department of Physics, Centre for Nonlinear Studies and Beijing-Hong Kong-Singapore Joint Centre for Nonlinear and Complex Systems (Hong Kong), Institute of Computational and Theoretical Studies, Hong Kong Baptist University, Kowloon Tong, Hong Kong
2: Department of Psychology, Carl von Ossietzky Universität Oldenburg, 26129 Oldenburg, Germany
3: Shanghai Key Laboratory of Brain Functional Genomics (Ministry of Education), School of Psychology and Cognitive Science, East China Normal University, 200062 Shanghai, China
4: Department of Psychology, Humboldt University at Berlin, Berlin, Germany
5: Department of Psychology, Zhejiang Normal University, Jin Hua, China
6: Research Center Neurosensory Science, Carl von Ossietzky Universität Oldenburg, Germany
7: Department of Physics, Zhejiang University, 310000 Hangzhou, China

*Corresponding authors. # Senior authors

E-mail addresses: andrea.hildebrandt@uni-oldenburg.de (A. Hildebrandt), cszhou@hkbu.edu.hk (C. Zhou).

**Abstract**

Over the last decades, cognitive psychology has come to fair consensus about the ontological structure of human intelligence. However, it remains an open question, whether anatomical properties of the brain support the same ontology. The present study explored the ontological structure derived from neuroanatomical networks associated with performance in 15 cognitive tasks indicating various abilities. Results suggest that the brain-derived (neurometric) ontology is partly consistent with the cognitive performance-derived (psychometric) ontology. However, there are interpretable and complementary differences as well. Moreover, the cortical areas associated with different inferred abilities are segregated, with little or no overlap. Nevertheless, these spatially segregated cortical areas are integrated via denser white matter structural connections as compared with the general brain connectome. The integration of ability-related cortical networks constitutes a neural counterpart to the psychometric construct of general intelligence, while the consistency and differences between psychometric and neurometric ontologies represent crucial pieces of knowledge for theory building, clinical diagnostics and treatment.

**Keywords**: Cognitive ontology, individual differences, neuroanatomical properties of the brain, brain networks, segregation and integration

**Introduction**

Broadly accepted cognitive ontologies are crucial for unified neurocognitive theories, diagnostics and mental health therapy. Over many decades, research on human cognition and intelligence, which aims at describing, measuring, and classifying abilities and unveiling their ontological structure, has been exclusively based on covariations of inter-individual performances across multiple tasks. Since the unified theory of cognitive abilities, known as the Cattell–Horn–Carroll (CHC) model, was compiled and disseminated (McGrew, 2009), there has been widespread consensus on the psychometric ontology of intelligence, notwithstanding some caveats (Schulze, 2005). According to the CHC model, stratum I comprises individual differences in a large number of cognitive tasks that require maximal performance in terms of speed or/and accuracy. Stratum II incorporates broad latent abilities that are not directly observable but task-derived, for example, fluid reasoning (*Gf*), comprehension knowledge (*Gc*, also called crystallized intelligence), short-term memory (*Gsm*), long-term storage and retrieval (*Glr*), and cognitive processing speed (*Gs*). These broad but domain-specific abilities are nevertheless positively associated with each other. This positive manifold is accounted for in the CHC model by a general factor of intelligence (*g*) at stratum III. A somewhat separate literature on ontological entities of human intelligence concerns executive functions (*EF*). The most widely accepted suggestions about the ontological entities of *EF* encompass abilities of working memory, shifting, and inhibition (e.g., Miyake et al., 2000).

Cognitive neuroscience – a much younger discipline than the psychometrics of intelligence – investigates, among others, the anatomical and functional properties and neural mechanisms underlying the ontological entities of human cognition. Early studies focused on stratum I abilities, aiming to understand the neuroanatomical basis and neurofunctional mechanisms underpinning the mastery of specific cognitive tasks. In this line, studies adopted connectome-based predictive modeling (e.g., Finn et al., 2015; Shen et al., 2017), and mapped brain properties and performance scores in specific cognitive tasks (e.g., Beaty et al., 2014; Cui et al., 2018; de Mooij et al., 2018; Kristanto et al., 2020). As a further step toward generalization, latent cognitive abilities at stratum II of the CHC model were mapped onto brain structure and function (e.g., Barbey, 2018; Colom et al., 2006; Dubois et al., 2018; Jung & Haier, 2007; Kovacs & Conway, 2016; M. Liu et al., 2020). The latter studies generated neuroanatomical and neurofunctional explanations of the psychometric ontological entities of cognition beyond specific tasks. In a further step, the quest for brain-derived ontological entities of cognition has recently gained traction. In this perspective, the psychometric entities are not taken for granted but are explored in a more bottom-up fashion (Poldrack et al., 2011) .

In the attempt to derive a brain-based ontology in addition to the established psychometric ontology, the cognitive neuroscience community has insufficiently appreciated the psychometric desiderata of distinguishing between constructs and their operational measures (Cronbach & Meehl, 1955). That is, a single task, brain measure, or indicator contains idiosyncratic, specific characteristics, whereas generalizable ontological entities of cognition must be determined by multivariate analyses going beyond single tasks, measures, or indicators (Cocchi et al., 2013). Consequently, modern cognitive neuroscientific approaches to derive ontological entities must use multiple tasks to explore distinct indicators of anatomical and functional properties of the brain to support the existence of stratum II ontological entities of cognition (the top-down approach), and mine the brain to derive ontological entities that do not take psychometric entities for granted (the bottom-up approach, Lenartowicz et al., 2010).

The recent availability of large-scale, cognitive neuroscientific databases enables a bottom-up approach, allowing to explore ontological entities of human cognition on the basis of brain properties (King et al., 2019; Poldrack & Yarkoni, 2016). For example, Bolt and colleagues investigated which cognitive entities are revealed by brain activation patterns when participants perform a battery of cognitive tasks (Bolt et al., 2017). Unfortunately, the authors did not systematically explore the consistency and disparity between their derived neurometric ontology and the ontology as established in psychometrics. We argue that such descriptive comparison is necessary in order to substantiate the ontology of human cognition, which should accommodate the covariance structures of both task performance and the brain systems supporting these abilities (Anderson, 2015; Jonikaitis & Moore, 2019; Lenartowicz et al., 2010).

The present study contributes to establishing a formal cognitive ontology for psychology, cognitive neuroscience, and their applied disciplines by adopting a bottom-up approach (see Lenartowicz et al., 2010). We explored the cognitive ontology as derived from anatomical cortical properties frequently employed in brain-behavior studies (Bayard et al., 2020; Reese McKay et al., 2013; Tadayon et al., 2020; Williamson & Lyons, 2018). We refer to this as neurometric ontology. The brain properties analyzed in relation with performance in 15 psychometric tasks were cortical thickness, myelination, curvature, and sulcus depth. Importantly, we performed the analyses on a large dataset with resampling realizations to derive a robust neurometric ontology. Going beyond existing approaches we compared the neurometric ontology with the psychometric one revealed by covariances of performance measures in the same tasks. Specifically, we investigated the extent to which the derived neurometric ontology is consistent with or distinguishable from the established psychometric ontology at stratum II of the CHC model. We expected that the ability-related cortical areas are segregated, supporting the differentiation between the inferred cognitive ontological entities. Finally, we investigated how ability-related cortical areas are coupled via fiber connections to possibly substantiate the positive manifold of domain-specific psychometric entities underlying general intelligence at stratum III of the CHC model.

**Results**

Our general methodological framework is illustrated in Fig. 1, consisting of the exploration of neurometric ontology in several resampling realizations and the comparison of robust neurometric ontology with the psychometric ontology. We also performed further analyses on cortical areas related to the neurometric ontological entities to investigate the underlying anatomical network. The Methods section provides a detailed description of the applied analyses framework.

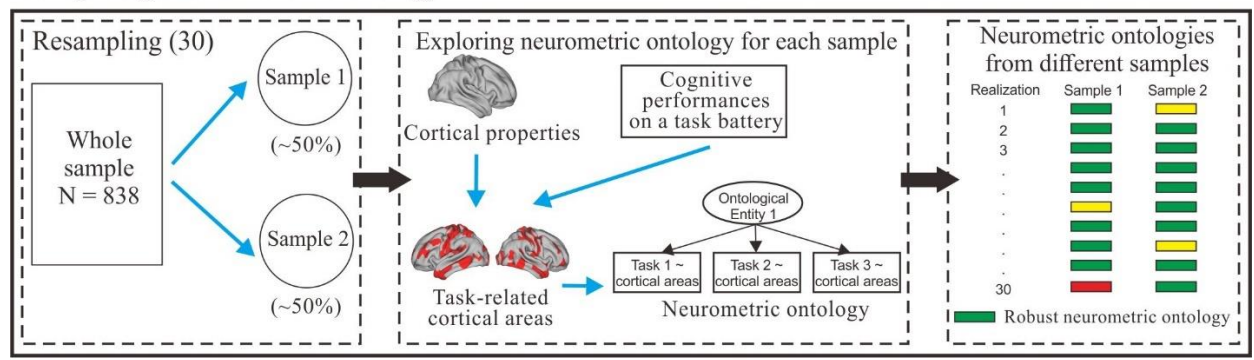

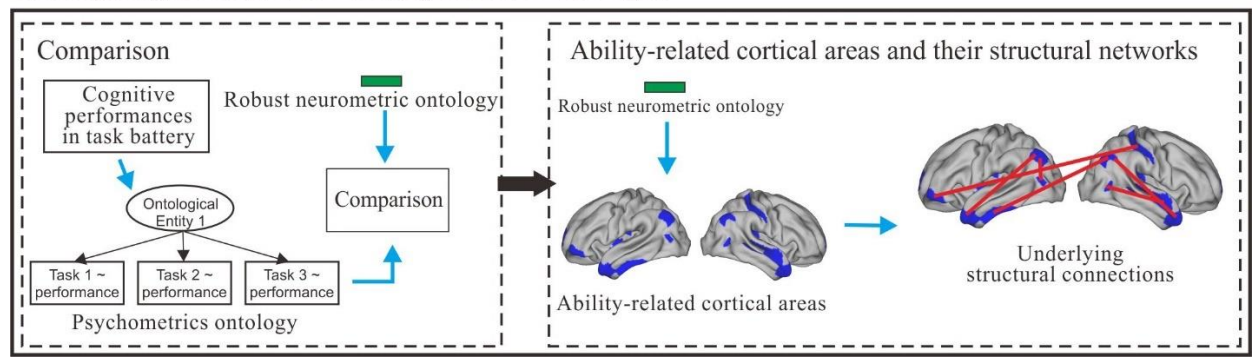

**Fig. 1. Framework of the study.** Summary of our methodological framework illustrating the systematic procedures. (A) To explore the neurometric ontology, we resampled the whole dataset to obtain several realizations, where each realization consisted of two independent samples. We identified task-related brain areas based on linear associations between brain properties (thickness, myelination, curvature, and sulcus depth) and task performance across individuals. Next, we quantified the overlap between task-related brain areas with these different anatomical properties by creating matrices of the Intersection Over Union (IOU) index. These matrices were then subjected to exploratory factor analysis (EFA) to obtain a neurometric ontology. Across samples and iterations, we selected the most robust, that is, most frequent neurometric ontology. (B) The most robust neurometric ontology was descriptively compared with the psychometric ontology derived from covariances across the same cognitive tasks estimated in the whole sample. To this aim, we applied confirmatory factor analysis (CFA) on the correlation matrix of task performance scores following the propositions of the CHC structure (Table 1). Next, we identified overlapping brain areas associated with the different entities of the neurometric ontology (henceforth referred to as ability-related brain areas) and investigated the underlying anatomical network connections between those areas.

**Neurometric versus psychometric ontological entities.** The psychometric ontology for descriptive reference was obtained by submitting the correlation matrix of task performance scores (Fig. 2A) to confirmatory factor analysis (CFA). For deriving the neurometric ontology, we applied Exploratory Factor Analysis (EFA) on the task-related cortical areas shared by each pairwise tasks. Shared cortical areas were quantified by the Intersection Over Union (IOU) index (Fig. 2B and 2C).

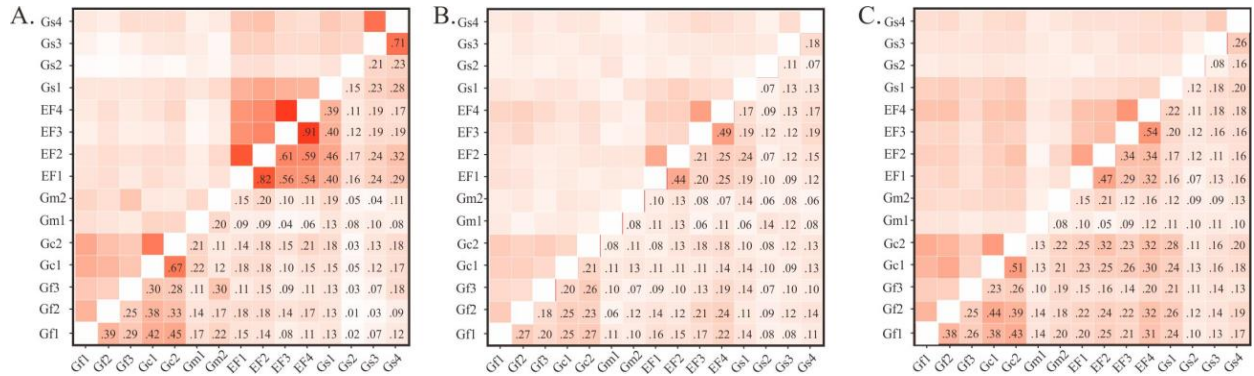

**Fig. 2. Correlation and IOU matrices.** (A) Correlation matrix of task performance scores in the whole sample; (B) Intersection Over Union (IOU) matrix from one random sample obtained by the resampling procedure; (C) IOU matrix obtained for the whole sample.

The psychometric ontology was obtained by a confirmatory model in line with the CHC theory. Following previous analyses of the HCP data for the same 15 tasks (Dubois et al., 2018; M. Liu et al., 2020), we estimated five domain-specific abilities as psychometric ontological entities—namely, *reasoning*, *comprehension knowledge*, *memory*, *executive functions*, and *mental speed*—corresponding to those listed in Table 1 (see *SI Appendix* Table S1 for the detailed task description), and depicted in Fig. 3A. Domain-specific ability factors were allowed to correlate with each other. To account for task specificity, three residual correlations between two pairs of *EF* tasks and one pair of *Gs* tasks were required. These indicators reflect two different conditions of the same tasks, implicating task specificity by design. The five correlated factors model fitted the matrix of task performance associations well (see *SI Appendix* Table S2, providing the statistical evaluation of model fit). The standardized factor loadings for all domain-specific abilities were significant, but their magnitude varied across ontological entities (see *SI Appendix* Table S3). The correlations between domain-specific abilities ranged from .26 to .79, with the smallest correlation observed between *EF* and *Gc* and the largest between *Gf* and *Gc* and between *EF* and *Gs*. This pattern of correlations between psychometric ontological entities is consistent with the vast literature on human intelligence (Carroll, 1993).

Aiming to derive a robust neurometric ontology, the analyses were repeated on 60 subsamples originating from 30 resampling realizations of two subsamples obtained from the whole dataset by random allocation. The results of the neurometric ontology across subsamples are listed in *SI Appendix* Table S4. Most frequently we observed a 3-factorial structure (47 out of 60 subsamples). In 19 out of 30 resampling realizations this structure was obtained in both samples. Importantly, the most frequently observed neurometric ontology was consistent with the one derived from the whole sample which reflected a 3-factorial structure (see *SI Appendix* Table S2 for the statistical evaluation of the model fit and Fig. S1 for the scree plot analysis to determine the number of factors).

Fig. 3B illustrates a simplified representation of the neurometric ontology obtained by resampling. We only depict loadings which were consistent across samples; *Gf1, Gf2, Gf3, Gc1, Gc2* were consistently assigned to factor 1; *EF1, EF2, EF3, EF4* were consistently assigned to factor 2; and *Gs1, Gs2, Gs3, Gs4* were consistently assigned to factor 3. However, since *Gm1* and *Gm2* were associated with different factors across samples, we did not assign them to any of the factors in the simplified structure. In addition, the factor loadings of *Gm1* and *Gm2* were smaller as compared to the other tasks within the same factor (refer to *SI Appendix* Table S4 for details and Table S5 for an example of the loadings in one of the samples). Fig. 3C illustrates the simplified neurometric ontology derived from the whole sample. The entire factor loading matrix estimated by the EFA on the whole sample IOU matrix is displayed in *SI Appendix* Table S6. Not obviously interpretable loadings are shown as broken lines in Fig. 3C. As indicated in *SI Appendix* Table S6, these loadings are much weaker than those indicated by solid lines. Overall, both loading patterns obtained by resampling and in the whole sample suggest a first factor which is interpretable as *reasoning and comprehension knowledge* (*G*). This reflects shared brain properties associated with rather difficult tasks, scored by performance accuracy (covering *Gf1*, *Gf2*, *Gf3*, *Gc1*, and *Gc2*). A second factor (*EF*) reflects shared brain properties relevant for executive functions covering *EF1*, *EF2*, *EF3*, and *EF4*. Finally, a third factor (*Gs*) comprises mental speed-related brain areas shared across easy cognitive tasks scored by the swiftness of responses, covering *Gs1*, *Gs2*, *Gs3*, and *Gs4*.

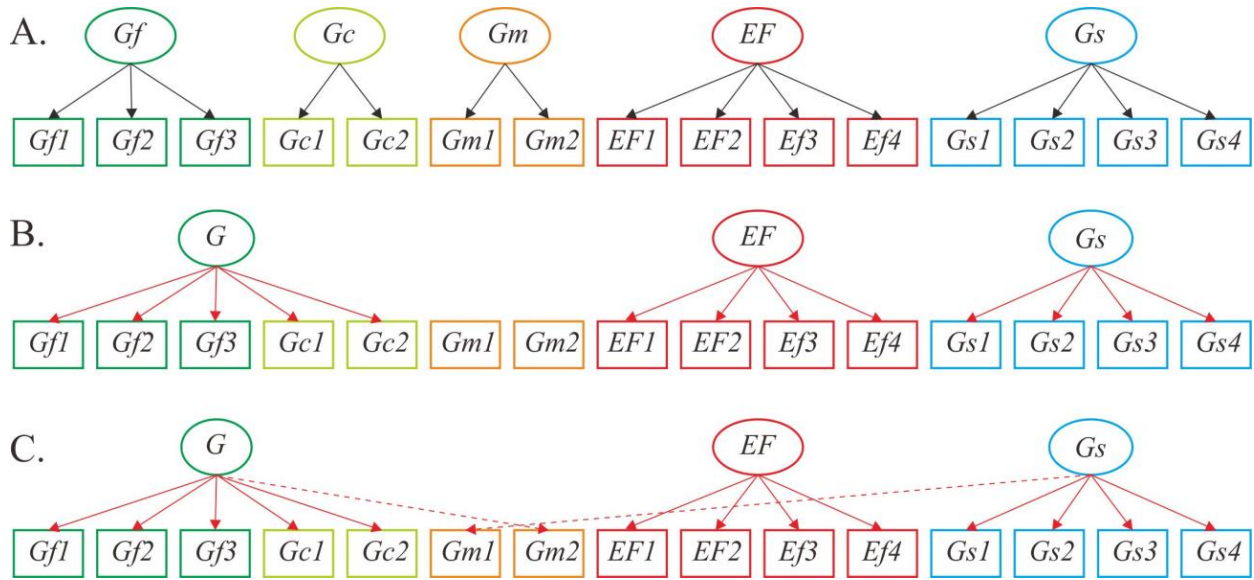

**Fig. 3. Schematic representation of the neurometric and psychometric ontology.** Descriptive comparison of the psychometric and neurometric cognitive ontology. (A) Psychometric ontology obtained by modeling the covariance structure of performance scores in 15 tasks; (B) The most reliable and robust neurometric ontology obtained by resampling; (C) Neurometric ontology derived from the Intersection Over Union (IOU) matrix of task-related brain areas in the whole sample. Note: In all analyses the factors were correlated but the correlations are not displayed here for simplicity. Details on model fit and loading estimates are provided in *SI Appendix* Tables S2–6. The solid lines indicate the ontological structure used for further analysis, whereas the broken lines indicate non-interpretable, weak loadings.

By comparison, the neurometric ontology derived from the IOU matrices of task-related brain areas (both from resampling and the whole sample) revealed a partly consistent, but less differentiated ontological structure as compared with the established psychometric ontology. To further explore, we fitted a model mimicking the cognitive ontological entities to the whole sample's IOU matrix using CFA. As summarized in *SI Appendix* Table S2, the fit of this model was poor. Furthermore, the model parameter estimates indicated that the IOU matrix did not reflect a 5-factor structure. This was mainly because in the IOUs, *Gf*, *Gc*, and *Gm* were highly correlated and not differentiable. Thus, these analyses further confirmed that task-related cortical areas reveal substantially different ontological entities as compared to those derived from associations of task performance scores across individuals.

**Cortical areas corresponding to the neurometric ontological entities.** We further identified the cortical areas corresponding to the neurometric entities. These areas were defined as the overlapping task-related cortical areas subsumed under a given entity in the EFA. For instance, the entity *Gs* derived from the IOU matrix (Fig. 3B) is reflected by cortical properties associated with tasks *Gs1*, *Gs2*, *Gs3*, and *Gs4*. Thus, the cortical areas corresponding to *Gs* are those that are shared by the tasks *Gs1*, *Gs2*, *Gs3*, and *Gs4*. Note that in the following, the 3-factorial structure supported by the whole sample's data will serve as reference neurometric ontology. Thus, ability-related cortical areas used task-related cortical areas as the basis, which were identified from the correlation between anatomical brain properties and performance scores from all participants (Fig. S3).

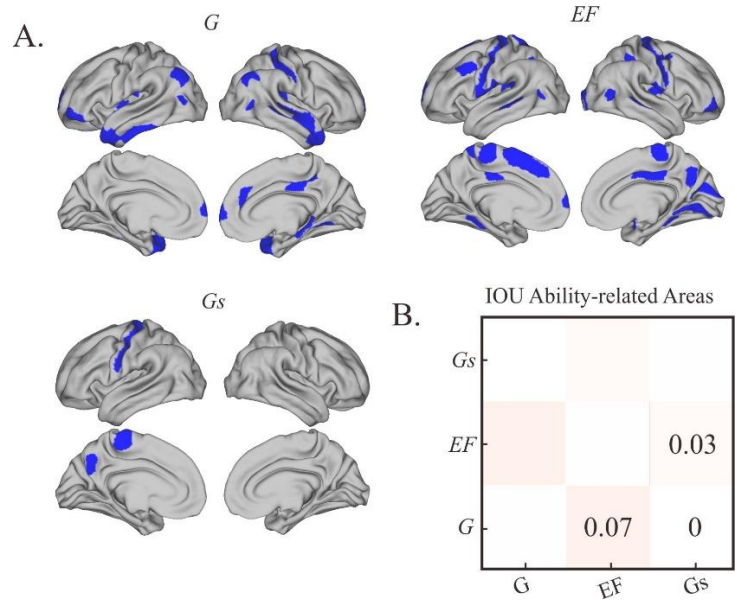

**Fig. 4. Inferred ability-related cortical areas. (A)** Brain areas corresponding to the neurometric ontological entities; **(B)** Pairwise overlaps computed from the IOU index of the ontological-entity-related brain areas reflecting domain specific abilities.

Fig. 4A illustrates the ability-related cortical areas for all the three neurometric ontological entities. In details, *G, EF,* and *Gs* respectively involved 26, 32, and 2 cortical areas, respectively. The areas associated with the *G* factor were dominated by frontal, lateral temporal, and auditory association cortex. *EF* was neuro-anatomically represented by more widely distributed cortical networks from both hemispheres, covering the prefrontal, motor, parietal, and the visual cortex. Finally, there were only two areas identified for *Gs*: area 4 in the motor cortex and area 7m in the cingulate cortex.

As the ability domain specific areas seem to be segregated across ontological entities, we quantified their exclusiveness by computing their pairwise overlap using the IOU index (see Methods), as shown in Fig. 4B. We found little to no overlap between the ability-related cortical areas. Specifically, the overlap was small between *G* and *EF* (.07), and between *EF* and *Gs* (.03). For reference, the overlaps of task-related cortical areas between tasks *EF1-EF2* and *EF3-EF4*, which were corresponding to the same ability (*EF*), were .47 and .54, respectively (see Fig. 2A). Interestingly, *G* and *Gs* did not share any brain areas. These results show that the ability-related cortical areas were well segregated among the entities of the neurometric ontology.

**Anatomical connectivity between ability-related cortical areas**. In the ontology of human intelligence, the positive manifold among abilities is accounted for by general intelligence at stratum III. Hence we asked how the segregated ability-related cortical areas might support such a positive manifold. Thus, we analyzed the network connectivity underlying the segregated ability-related cortical areas. Structural connectivity was obtained by probabilistic tractography averaged across individuals (see Methods). The ability-related cortical areas as defined above for each ontological entity were set as core areas, whereas the union of all task-related cortical areas of the corresponding ontological entity, excluding the core, were considered as extended areas. We examined (i) the connections between the core areas, and (ii) the connections between the extended areas. Both connectivity levels can be applied within areas of the same ontological entity and between entities.

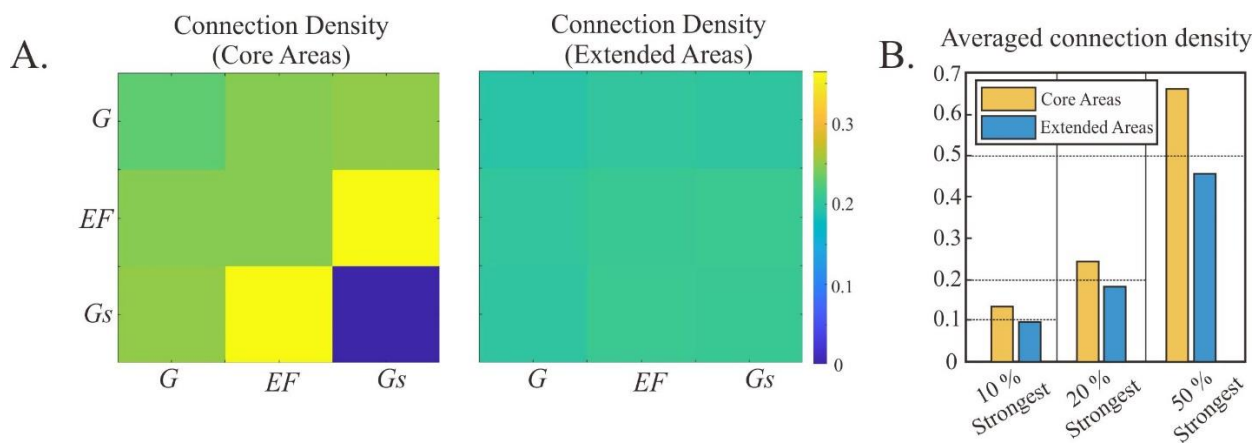

**Fig. 5. Comparison of the anatomical connection density between the ability-related cortical areas (core areas) and the other task-related areas (extended areas).** (A) Connection density within and between cognitive entities included in the neurometric ontology at network density threshold of 20%; (B) The average connection density of core areas across different entities and all extended areas at different network density thresholds (10%, 20%, and 50%, horizontal dotted lines) in the whole structural connectivity network.

We applied network density thresholds of 10%, 20%, and 50% of the strongest connections in the weighted group-averaged connectivity matrix to identify existing connections (binary 0 or 1). For each level of connections (core and extended areas), we obtained the connection density as the ratio between the number of existing connections and the number of possible connections. Under all thresholds, the connection density of the core areas was clearly larger and that of extended areas was clearly smaller than the average connection density of the whole brain network (Fig. 5B). This finding indicates that, although the ability-related cortical areas were segregated in their regional anatomical properties, they were nevertheless densely connected by subcortical white matter projections, such that their average connection density is greater than that of the brain connectome. An exception holds for *Gs* at connectivity thresholds 10% and 20%, where the core areas were not connected (Fig. 5A). This might be due to the fact that there were only 2 brain areas related to *Gs*, considerably fewer than for the other abilities. However, at the 50% connectivity threshold (see *SI Appendix* Fig. S2), the areas related to *Gs* were structurally connected. It indicated that the areas for Gs were indeed connected by white matter fibers.

**Discussion**

We explored the ontology reflected in the task-related anatomical properties of the cortex (neurometric ontology) and compared it with the widely established psychometric ontology based on task performance correlations alone as reflected in the CHC model. We investigated anatomical properties that were frequently used as biomarkers of cognition (Bayard et al., 2020). Specifically, we selected cortical thickness as a known neural property associated with crystallized intelligence (Tadayon et al., 2020) and myelination which was shown to be related with neuronal circuit plasticity (Williamson & Lyons, 2018). Finally, cortical curvature and sulcus depth have been found to be genetically mediated and associated with cognition and neurodevelopmental disorders (i.e., Tourette syndrome; Müllner et al., 2015; Reese McKay et al., 2013). In order to cross-validate the neurometric ontology, we used a resampling approach to create 60 samples and explored the neurometric ontology based on each sample (see Methods). We found that the most frequently observed ontological structure implies three entities: *G*, *EF*, and *Gs*. The neurometric ontology was partly consistent with the psychometric ontology, but it was less differentiated. Interestingly, the ability-related brain areas demonstrated competing features: On the one hand, the cortical areas associated with different abilities were spatially segregated, as there was little overlap. On the other hand, they were also integrated via structural fiber connections that were denser than the average brain connectome. Next, we discuss the three main findings of the current study.

**Consistencies and differences between the psychometric and neurometric cognitive ontology**.

The neurometric ontology was derived from the IOU matrix quantifying overlapping task-related cortical areas identified from correlations between anatomical brain properties and task performance. The psychometric ontology was derived from correlations of task performance scores. Here, the question may arise whether the IOU and task performance correlation matrices are similar in structure just because they both involve task performance and hence are redundant. We argue that correlation between cognitive task performance scores may be observed even for tasks instantiated by distinct brain networks (Oberauer et al., 2005). Thus, correlations between task performances are not necessarily due to a greater neuroanatomical overlap between task associated brain areas and hence higher values in the IOU index. For example, the tasks *Gf3* and *Gm2* were correlated in performance .3, but their IOU index was .1 (Fig. 2). In contrast, *EF4* and *Gc1* showed a small correlation (.15), but the corresponding brain area overlap in the IOU index was .3. Hence, the neurometric and psychometric ontology might in principle converge or dissociate.

Indeed, we found that the neurometric ontology aligns with the psychometric ontology to some extent, but that they also dissociate in other respect. The most evident consistency is that the *Gs* and *EF* entities were found in both. These similarities are remarkable, because the dependent measures subjected to factor analyses (performance speed and accuracy versus anatomical properties of the brain) are very different in nature. It can thus be concluded that there is a partially robust isomorphy between cognitive ontological entities derived from cognitive behavior and their associated neuroanatomical properties. This validates important aspects of the CHC structure of human cognition from a neurometric point of view.

In contrast to these consistencies, we also observed interesting and interpretable differences between the psychometric and neurometric ontology. First, the IOU matrix did not show a separation between *Gf* and *Gc*. We thus termed the overarching ontological entity as *reasoning and comprehension knowledge* (*G*). This unified *Gf* and *Gc* is in line with Cattell's view on general intelligence, encompassing these two facets of cognition (Cattell, 1943). A strong relationship between *Gf* and *Gc* was also suggested in terms of skill acquisition (Wenger et al., 2017) and emotional intelligence (Olderbak et al., 2019).

Moreover, taking a closer look at the ontological entities derived from resampling, the second-most frequently obtained neurometric ontology was a 4-factorial structure, in which the separation of *EF* was evident (i.e., *EF1* and *EF2* were assigned to a single factor, while *EF3* and *EF4* were assigned to another factor). This observation can be explained by referring to Table S1: *EF1* and *EF2* tapped into *switching*, while *EF3* and *EF4* are measuring *inhibition*. It is interesting that switching and inhibition, which are often treated as one overarching entity (*EF*) in psychometrics (for an exception see Miyake et al., 2000), can be distinguished when analyzing the structure of overlapping cortical areas.

Another difference between the psychometric and neurometric ontology revealed that *Gm* could not be identified as an independent cognitive entity based on task-related cortical areas. Note that cortical areas related to memory tasks were scattered into several neurometric ontological entities (i.e., *G* and *Gs* in case of the IOU matrix from the whole sample, see Fig. 3C). This may indicate that cortical areas associated with *Gm* are complex and widely distributed, such that they may not be quantifiable by neuroanatomical properties of cortical regions alone (Nadel & Hardt, 2011; Squire, 2009). This is illustrated by Fig. S3, where the areas related to *Gm1* and *Gm2* were, although fewer compared to other tasks, distributed across the cortex. However, detailed analyses of task-related cortical areas in terms of these properties may facilitate the understanding of brain networks underlying *Gm* in the future. Here we observed overlaps with the brain networks associated with *G* and *Gs*. In addition, *Gm* was also found to be associated with anatomical properties of subcortical areas, which were not well captured by the MMP atlas (Koshiyama et al., 2018; Lee et al., 2019) applied in this study.

We argue that the results above, indicating similarity and explainable dissimilarity between psychometric and neurometric cognitive ontologies are compelling, considering that we explored only a limited number of anatomical properties of cortical regions. These findings have the potential to prompt further psycho-neuro-informatics studies to explore the neurometric ontology using a broader range of anatomical, biochemical, genomic and functional properties of the brain supported by different computational modeling approaches (Guest & Martin, 2021). From a more general perspective, we emphasize that the present comparative approach, focusing on similarities and dissimilarities, is highly valuable for exploring whether ontological entities derived from psychometrics (stratum II in the CHC model) translate into brain anatomy and function (Anderson, 2015). The potential translation is not only of theoretical importance, but also relevant for neuropsychological diagnostics and therapy.

**Segregation of ability-related cortical areas in the neurometric ontology.** After analyzing similarities of and differences between the psychometric and neurometric ontologies, we investigated whether the entities of the neurometric ontology are built upon distinct neural networks. This approach was inspired by a study on cognitive control (i.e., executive functions) by Lenartowicz and colleagues (Lenartowicz et al., 2010). The authors applied a machine learning approach to classify data from more than one hundred neuroimaging studies, aiming to isolate component operations with similar and dissimilar brain-activation patterns across experiments (Lenartowicz et al., 2010). They found that ontological entities described in the psychometric literature of cognitive control were generally also differentiable on the basis of brain activation patterns. However, while *shifting* was consistently identified as a specific ontological entity in psychometrics, in the brain-activity covariance structure it could not be discriminated from *response selection* or *response inhibition* (Miyake et al., 2000).

Here we compare the cortical areas corresponding to the entities in the neurometric ontology with the results of other studies on ability-related anatomical brain networks. First, *G* was dominated by frontal, medial temporal, and auditory association cortices. This agrees with a previous study claiming that the cortical thickness of frontal and parietal cortex is related to intelligence in general (IQ; Bajaj et al., 2018). In addition, the association between temporal and auditory cortex with fluid and crystallized intelligence, as part of *G* in the present study, has been studied using structural MRI (Kristanto et al., 2020; Phinney et al., 2007). Areas related to *EF* were widely distributed across the cortical surface, suggesting that *EF* is a complex ability that requires integration of different brain networks, in line with previous studies on *EF* (Niendam et al., 2012; Zink et al., 2021). For *Gs*, our results indicate that motor cortex is relevant to *mental speed*. Considering that the present study exclusively considered anatomical properties, our finding complements previous research on the association between the motor cortex and mental speed, mostly found in functional properties. For example, beta-band oscillations in the motor cortex have been related to movement speed in human and macaque (Khanna & Carmena, 2017; Zhang et al., 2020). Another study showed that the functional connection between motor and visual areas were predictive for processing speed (Gao et al., 2020). Moreover, aside from their consistency with some previous studies, we found that the areas corresponding to each entity had little overlap with other entities or even showed complete exclusiveness. Although there are some small overlaps of *EF* and other entities, *G* and *Gs* do not share any cortical areas. This is in line with research on speed and accuracy, suggesting that these performance outcomes are substantiated by independent brain networks. For example, using EEG, Perri et al. (2014) found that accuracy tasks trigger activity in the frontal cortex, while speed tasks rely on activity in the motor cortex.

The present findings about the neurometric ontology, namely (i) its consistency with and interpretable dissimilarities to the psychometric ontology, in terms of inferred cognitive entities, and (ii) the exclusive and functionally relevant brain areas corresponding to the inferred entities (see previous sections) suggest that the neurometric ontology may be an important complement for psychometric entities of human cognition. There are at least three main reasons for this claim. First, convergence or divergence between the psychometric and neurometric cognitive ontology helps to identify and resolve weaknesses reflected in psychological terminology and theory building (Anderson, 2015; Poldrack & Yarkoni, 2016). Second, it will elevate the quest for brain behavior relationships from the task-specific level to the theoretically more interesting level of domain-specific abilities. Third, the juxtaposition of the psychometric and neurometric ontology is relevant from an applied clinical perspective, for which the mapping of psychological functions to their corresponding brain systems is essential (Fornito et al., 2017).

**Dense fibre connections between segregated ability-related cortical areas.** The segregated core areas for different neurometric entities, especially at 50% connectivity thresholds (see Fig. 5 and *SI Appendix* Fig. S2), were found to be more densely connected, whereas the extended areas of different entities were more sparsely connected than the average whole brain connectome. Studies of segregated modules or communities in structural or functional brain networks typically find inter-module connectivity to be sparser than intra-module connectivity (Meunier et al., 2009, 2010). Thus, our finding of denser connections between segregated core areas of different abilities is unexpected, and to the best of our knowledge unparalleled in the literature. Our findings indicate that segregated ability-related cortical areas are actually integrated in terms of structural connections.

We therefore want to emphasize two important but at first sight contradictory features of the inferred ability-related brain areas: On the one hand, they are exclusive (segregated), but on the other hand more densely connected (integrated by fibre tracts) than the average of the brain. We interpret the "segregation" feature to reflect the partial independence of stratum II entities in the psychometric structure of cognition. Furthermore, the relatively dense connections of the ability-related brain areas across ontological entities, i.e., the "integration" feature, appears to support the positive manifold of the abilities that produces general intelligence at stratum III in the CHC ontology. This interpretation offers a new perspective on how the structure of human intelligence can be understood in terms of underlying neural clusters in the regional neuroanatomical properties and fiber network projections in the brain to support both segregation and integration. The segregation and integration configuration of brain networks have been studied to achieve a better understanding of cognitive abilities, including their relationships with general intelligence (Bassett & Bullmore, 2017; Deco et al., 2015; Fair et al., 2007; Oldham & Fornito, 2019; Sporns, 2013; Wang et al., 2019, 2021). However, these previous studies focused on functional properties of the brain, prompting our question whether also neuroanatomical properties reveal similar segregation and integration organizational features. The results of the present study provide an affirmative answer to this question. Future studies may specifically investigate the association between the structural connections and the inter-individual variability between segregated ability-related cortical areas and general intelligence.

*Limitations and outlooks*

The present study is not without limitations. First, it could only consider 15 cognitive tasks and focused on four anatomical properties of the brain. Applying a larger number of cognitive tasks and measuring more brain properties, such as cytoarchitecture, connectivity, biochemistry, and gene expression, is a promising direction in this line of research. In addition, behavioral outcomes may not rely only on regional properties, as studied here, but also on connection properties. By including the connection properties (which may change the modularity and integration), the brain areas which were not well separable in term of anatomical properties could become more differentiated. Recent network neuroscience investigates how cognitive functions are reflected in brain networks. However, most of these studies have only considered connection properties. Future studies should consider both regional and connection properties to explain the segregation and integration of task-related brain areas (e.g. Wang et al., 2019, 2021).

Moreover, we derived the ability-related brain areas by overlapping task-related brain areas across corresponding tasks of a given cognitive entity. Therefore, employing a larger number of tasks which are representative for the targeted abilities and providing more indicators should result in a more robust determination of the psychometric ontology. Furthermore, a particular cognitive task may not be sufficiently pure to measure a specific cognitive function (see Poldrack & Yarkoni, 2016). For example, the *Gs1* task challenges visual and semantic processing in addition to mental speed. Therefore, future work should rely on cognitive tasks that are as pure as possible, to capture specific cognitive functions.

Finally, the task-related cortical areas were identified by linear associations. As suggested in the literature (Kriegeskorte, 2015), a non-linear approach using, for instance, artificial neural networks could be applied, to afford a more precise mapping of anatomical properties of the brain and cognitive behavior.

*Conclusion*

We applied a novel approach to derive a neurometric cognitive ontology from anatomical properties of the brain, and compared this ontology with a largely standard psychometric ontology derived from task performance. The results revealed that the entities of the neurometric ontology partly reflect the current psychometric view. In important respects, the brain areas related to the neurometric ontologies correspond to findings from previous imaging studies of cognitive abilities. We also found sparse (or even no) overlap between brain areas related to different neurometric ontological entities, reflecting the neural reality of separable psychometric cognitive entities. Although the cognitive entities were segregated, the structural connections between their core brain areas were relatively dense. These findings seem to explain the moderate correlations of psychometric abilities at stratum II by the result of segregated ability-related cortical areas, and the positive manifold captured at stratum III by the revealed dense fiber connections between cortical core areas.

## Materials and Methods

### Materials

*Participants*

This study used the publicly available data from the WU-Minn Young Adult Human Connectome Project (HCP), covering several magnetic resonance imaging (MRI) modalities, including resting-state functional MRI, diffusion MRI, and structural MRI (Van Essen et al., 2013). The HCP database also provides demographic data and performance measures for a number of cognitive tasks. Of the 1,206 data sets in the HCP database at the time of study, those with incomplete information were excluded, yielding a final sample of $N = 838$ (449 females, 760 right-handed, and three ambidextrous) individuals. Age ranged from 22 to 35 years. Across all subjects, there were 401 different families identified by the ID of the mother and father of the participants. For more detailed information on data acquisition protocols of the HCP please refer to the project's website (https://www.humanconnectome.org/). The local ethics committee of Washington University approved the HCP study.

*Anatomical measures*

We considered four properties of the cortex, derived from structural MRI: *thickness*, *myelination*, *curvature*, and *sulcus depth*. The T1-weighted and T2-weighted images (see (Glasser & van Essen, 2011) for the details of the scanning processes) in the data base has been pre-processed to correct distortions and align the images to the Montreal Neurological Institute space template (Glasser et al., 2013). Using the pre-processed images, gray and white matter of the cortical surface was reconstructed. All anatomical properties of the cortex used in the present study were estimated from the surface reconstruction. Cortical thickness was measured as distance between the pial surface and the gray–white matter border (Alvarez et al., 2019). The thickness of the cortex correlates with the number of neurons (Herculano-Houzel et al., 2013). Myelination, that is, myelin content of a given voxel, was determined as the ratio of T1 and T2 weights (Glasser & van Essen, 2011), with a higher ratios indicating the higher abundance of myelin in a given voxel. Curvature indicates the bending of each vertex of the cortical surface, and a higher curvature indicates sharper bending, with positive values corresponding to convex bending (Shimony et al., 2016). Lastly, sulcus depth was calculated as the distance between mid-surface gyri and sulci (Müllner et al., 2015). Curvature and sulcus depth contribute to determining the sulcal patterns predating the acquisition of cognitive skills (Mangin et al., 2010). All voxel-wise data, including both individual and group averages, were taken from the HCP website. We used the Multi-Modal Parcellation (MMP) atlas (Glasser et al., 2016), which divides each hemisphere of the brain into 180 areas. We averaged the voxel-wise property measures within a given brain area to obtain an area-specific measure of the respective property. Thus, for each participant, four brain properties were featured in each brain area.

For the analysis of connections among brain areas, we measured structural connectivity based on white matter projections between brain areas in the MMP atlas. We applied probabilistic tractography to the diffusion MRI data to trace white matter connections. Briefly, we set seed and target areas for a pair of brain areas. From each vertex of the seed area, we generated 5,000 streamlines and counted how many reached the target area. When the streamlines met with a voxel whose fractional anisotropy was less than .1, the propagation was stopped. There were two streamline directions, given that an area can serve as both, seed and target. Thus, the final structural connectivity between two areas was the average of two directional connectivity probabilities (see Liu et al., 2019 for details).

*Cognitive tasks*

Performance in 15 cognitive tasks was used for the present analyses. The tasks were adopted from the task-evoked fMRI measurement (in-scanner test), NIH Toolbox, and Penn Computerized Cognitive Battery (Barch et al., 2013; Dubois et al., 2018; Van Essen et al., 2013), and were selected by the HCP as a representative set of tasks covering most of the domain-specific abilities at stratum II, according to the CHC model (see above). Table 1 provides a summary of the tasks grouped into different stratum II abilities. We also included executive function (*EF)* as a domain-specific ability (Zink et al., 2021) because it plays a crucial role in human cognition and has pivotal diagnostic relevance in many mental conditions.

The indicators obtained in the tasks *Gf1*, *Gf2*, *Gf3*, *Gc1*, *Gc2*, *Gm1*, and *Gm2* were performance accuracies (Weintraub et al., 2013), whereas in the tasks *EF1*, *EF2*, *EF3*, *EF4*, *Gs1*, *Gs2*, *Gs3*, and *Gs4* we used inverted reaction times of correct responses, indicating the number of trials correctly solved per second. All performance indicators were standardized prior to statistical analyses. In all tasks, higher values correspond to better performance.

**Table 1. Cognitive tasks and associated domain-specific abilities included in the present study**

| Domain-specific abilities | Tasks |
| --- | --- |
| Reasoning *(Gf)* | Raven's Progressive Matrices *(Gf1)* |
| | Spatial Orientation Processing *(Gf2)* |
| | List-Sorting Working Memory *(Gf3)* |
| Comprehension Knowledge *(Gc)* | Oral Reading Recognition Test *(Gc1)* |
| | Vocabulary Comprehension *(Gc2)* |
| Memory *(Gm)* | Verbal Episodic Memory *(Gm1)* |
| | Picture Sequence Memory *(Gm2)* |
| Executive Function *(EF)* | Dimensional Change Card Sort – Color *(EF1)* |
|  | Dimensional Change Card Sort – Shape *(EF1)* |
|  | Flanker Inhibitory Control and Attention Task – Congruent *(EF3)* |
|  | Flanker Inhibitory Control and Attention Task – Incongruent *(EF4)* |
| Mental Speed *(Gs)* | Pattern Comparison Processing Speed *(Gs1)* |
|  | Sustained Attention *(Gs2)* |
|  | Relational Processing 1 *(Gs3)*\* |
|  | Relational Processing 2 *(Gs4)*\* |

\* Tasks were performed during the functional magnetic resonance imaging scanning (in-scanner tasks).

For a detailed description of these tasks see Table S1 and in the HCP manual (Van Essen et al., 2013).

**Methods**

*Resampling Analyses*

Aiming to obtain robust and reliable results, we analyzed subsets of data from several resampling realizations. As depicted in Fig. 1A (left panel), we first resampled the dataset to obtain 30 realizations, where each realization consisted of two different subsamples while assigning siblings in the dataset to different subsamples. Therefore, in total, we obtained 60 samples. Each sample was used separately in the analysis of the neurometric ontology.

*Identifying task-related cortical areas*

We identified task-related cortical areas for each of the 15 tasks by correlating anatomical properties of the 360 brain areas with task performance across individuals (Fig. 1A, middle panel). The CORR function in MATLAB (https://www.mathworks.com/help/stats/corr.html) was applied to determine the Pearson's correlation coefficients. This analysis resulted in a single correlation value for each cortical area for a given brain property and task. Thus, for all tasks and the four cortical properties together, there were 15 × 4 correlation values in each cortical area. The CORR function additionally provides *p*-values; thresholding at $p < .05$ was applied to select relevant cortical areas associated with a given task. Thus, cortical areas that were significantly correlated with performance were regarded as task-related. The selection of the *p*-value threshold ($p < .05$) was based on a previous study (Kristanto et al., 2020) where a smaller *p*-value threshold ($p < .01$ or $p < .005$) had resulted in the identification of cortical areas specific for a particular cognitive task. In the present study, we explored the relationships among cognitive tasks in terms of shared cortical areas. Thus, we selected a more liberal *p*-value threshold ($p < .05$) which captured not only the cortical areas specific for a particular task, but also cortical areas shared between tasks.

As data mining was applied to four anatomical properties of the brain, four sets of task-related cortical areas were obtained for each task. For a multimodal representation, we determined the union of these sets of task-related cortical areas across the four anatomical properties. Thus, to qualify as task-related, a given cortical area had to be significantly correlated with the task score ($p < .05$) in at least one of the four properties. The obtained cortical networks for each of the 15 tasks from all samples were subjected to further analysis.

Note that we performed the identification of task-related cortical areas in each sample and in the whole sample. Fig. S3 depicts the cortical areas related to each task identified from the whole sample, together with their hemisphere-specific distributions. The masks of the task-related cortical areas are also provided as supplementary materials in form of *.gii* files (gifti) by using the MMP atlas. Fig. S3 additionally illustrates the contributions of different cortical properties to the task-related cortical areas. Importantly, taken together, the task-related cortical areas covered approximately 80% of the entire cortex.

*Quantifying the overlap of task-related cortical areas*

The task-related cortical areas were not exclusive to individual tasks, and therefore overlapped across some tasks. Therefore we quantified the overlap for every pair of tasks by using the well-known index of Intersection Over Union (IOU) (Jaccard, 1912; Rezatofighi et al., 2019). The IOU index is widely used in measuring the similarity between sample sets. In our case, the IOU index of two tasks was calculated as the ratio between the number of common brain areas related to both tasks and the total number of distinct brain areas in each of the two tasks, and expressed as follows:

$$IOU = \frac{|A \cap B|}{|A \cup B|}.$$

where, A and B are sets of related cortical areas for Task 1 and Task 2, respectively. The computation of IOU indices of all 15 task pairs resulted in a 15 × 15 IOU matrix. As the task-related cortical areas were based on the correlations between performance and brain properties, we emphasize that the IOU matrices reflect the covariance structure of task pairs across the participants in terms of neuroanatomical properties. Fig. 2B and 2C visualized the IOU matrices of task-related cortical areas identified from one of the samples and the whole sample. respectively.

*Exploring the structure of IOU matrices*

We derived the neurometric ontology from the IOU matrix of task-related cortical areas for each of the 60 participant samples. The factorial structure was assessed by exploratory factor analysis (EFA), using the psych package (https://cran.r-project.org/web/packages/psych/index.html) in R. EFA is a data-driven approach that allows determining the number of factors needed to explain the structure of an association matrix, and the pattern of factor loadings. As different ontological entities indicated by factors in EFA are expected to be correlated, oblique (promax) rotation was applied to achieve a simple structure with interpretable factors. The number of required factors was found by scree plots, based on eigenvalues of a principal components and comparison with random eigenvalues (Fig. S1).

For EFAs, model fit was evaluated with the $\chi^2$ test statistic and RMSEA. In large samples, the $\chi^2$ test statistic is highly sensitive. Thus, alternative fit indices play an important role in model fit evaluation. The acceptable value for the RMSEA is < .08 (Hu & Bentler, 1999) .

*Comparison with the psychometric ontology*

We compared the most frequently observed neurometric ontology derived from the structure of IOU matrices with the psychometric ontology established by covariance analyses of task performance scores as a reference model. First, we performed a confirmatory factor analysis (CFA) of the correlation matrix of task performance scores of the 15 tasks to provide a baseline model of cognitive ontological entities. For CFA we used the lavaan package (https://lavaan.ugent.be) in R Software for Statistical Computing (https://www.r-project.org/). The CFA model included the domain-specific cognitive ontological entities listed in Table 1. Factors were standardized for identification purposes, such that all factor loadings could be freely estimated. Model fit of CFA was quantified by the $\chi^2$ goodness-of-fit index ($\chi^2$), the comparative fit index (CFI), the root-mean-squared error of approximation (RMSEA), and the standardized root-mean-squared residual (SRMR). Acceptable values for these indices are > .95 for the CFI and < .08 for both the RMSEA and SRMR (Hu & Bentler, 1999).

CFA and EFA rely on different assumptions, but may lead to consistent factorial structures when applied to the same data (Chen et al., 2015; Mian et al., 2012; Murray et al., 2019). The methodological choices above reflect the assumptions made. There is vast knowledge on which tasks are loaded onto which cognitive entity in the psychometric ontological structure. Furthermore, previous studies of the HCP data (e.g., Dubois et al., 2018; M. Liu et al., 2020) have confirmed the domain-specific cognitive entities considered in the current study. Thus, the factorial structure provided by the CFA of the covariance matrix of performance scores is not specific for the present analyses; rather, it is a descriptive reference for interpreting the ontological entities derived from brain properties. Furthermore, there is no prior knowledge on how the IOU matrices of brain areas related to the 15 specific cognitive tasks are clustered. Thus, by applying a CFA to the IOU matrices mimicking the psychometric ontological structure, we investigated whether task-related brain areas reveal exactly the same entities as obtained from performance. Hence, exploring the structure of the IOU matrices and identify potential differences between psychometric and neurometric ontological entities, an EFA approach appears to be reasonable.

**Acknowledgments**

This work was supported by the Hong Kong Research Grant Council (RGC) (HKBU12301019, HKBU12200620) and the Hong Kong Baptist University (HKBU) Research Committee Interdisciplinary Research Matching Scheme (IRMS/16-17/04, IRCMS/18-19/SCI01). This research was conducted using the resources of the High-Performance Computing Cluster Centre, HKBU, which receives funding from the RGC, the University Grants Committee of the HKSAR and HKBU

**Author Contributions**

DK: Conceptualization, methodology, software, formal analysis, writing-original draft, visualization. XL: resources, methodology, software, formal analysis. WS: Conceptualization, writing-review and editing, supervision. AH: Conceptualization, methodology, formal analysis, writing-review and editing, supervision. CZ: Conceptualization, formal analysis, writing-review and editing, supervision, funding acquisition.

**Data Availability**

The datasets that support the findings of this study are available at http://www.humanconnectome.org/study/hcp-young-adult.

The codes used in this study is available at https://github.com/kristantodan12/Mining-the-brain

**Supplementary Materials**

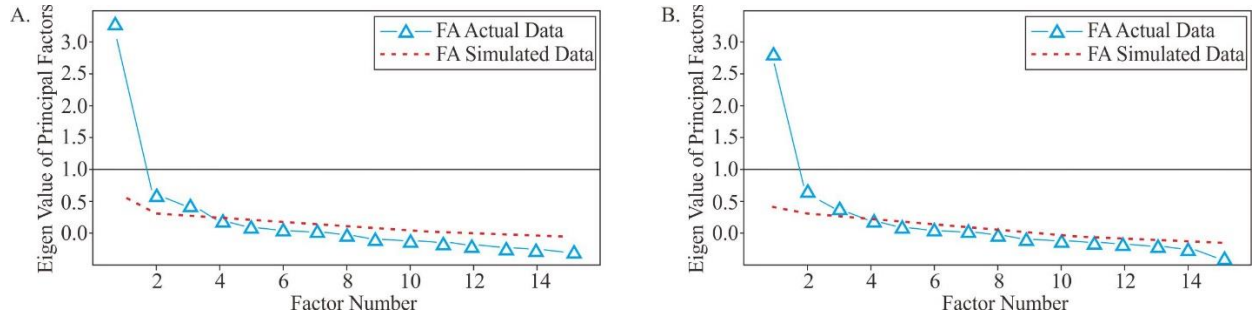

**Fig. S1. Scree plots of exploratory factor analyses of** IOU matrices. (A) A single sample from resampling (B) the whole sample. The x-axis represents the number of factors; the y-axis represents the eigenvalue of the principal factors from actual (blue line) and simulated data, expressed by random (reduced) correlation matrix (red dashed line). The suggested number of factors is determined by the number of eigenvalues above the points where the two lines intersect, which are 3 for both (A) and (B).

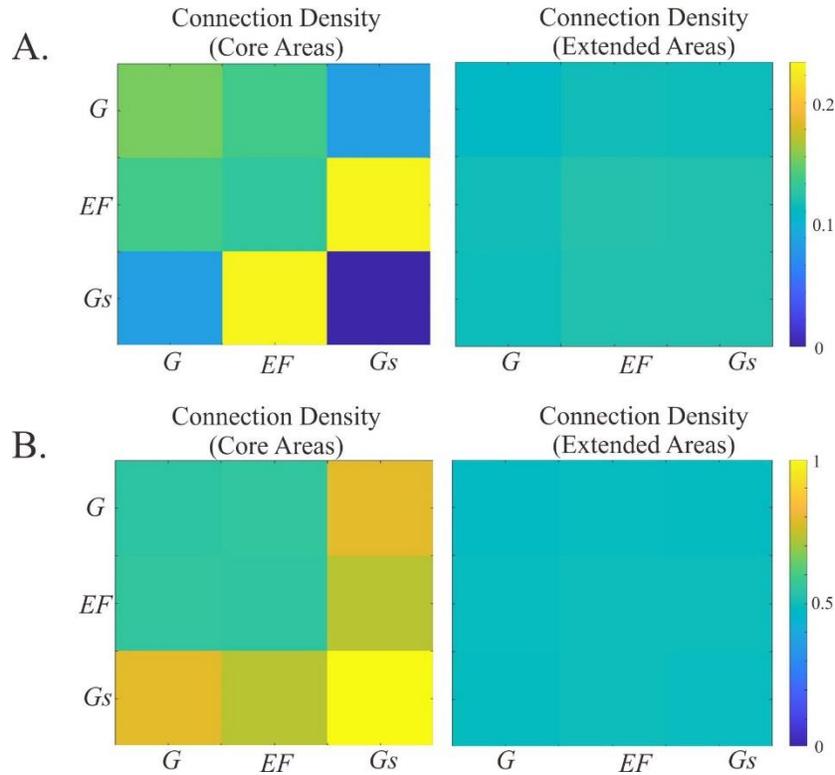

**Fig. S2 Anatomical connection of ability-related brain areas.** Comparison of the anatomical connection density between the ability-related brain areas (core areas) and the other task-related areas (extended areas). Connection density within and between cognitive entities included in the neurometric ontology at network density threshold of (A) 10% and (B) 50%.

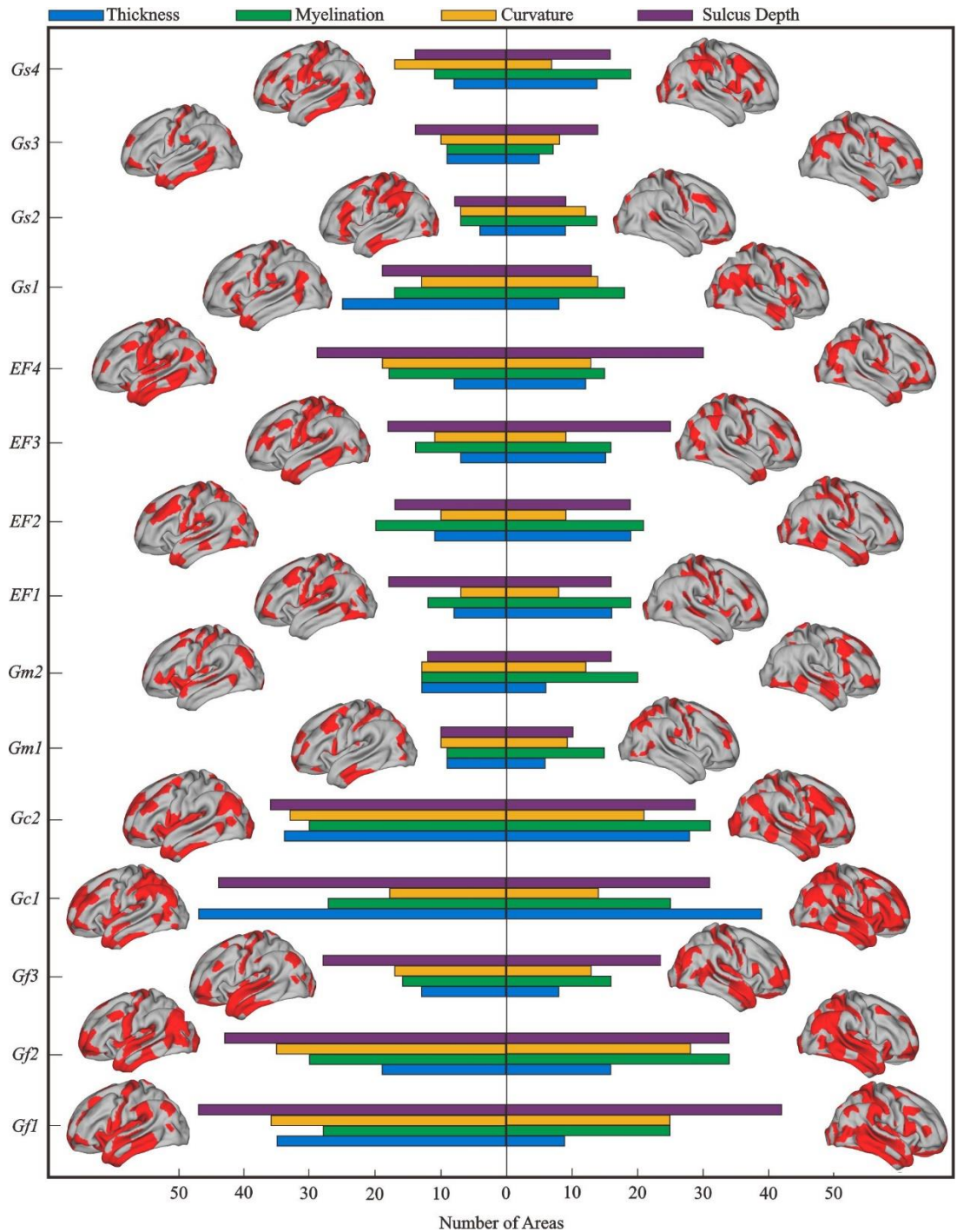

**Fig. S3. Distribution of task-related cortical areas.** Distribution of task-related cortical areas and the contribution of different anatomical brain properties in determining the task-related cortical areas in the left and right hemispheres. The areas on the brain maps (in red) represent the task-related brain areas from each task. Note that the task-related cortical areas are the union of the related areas identified from all anatomical brain properties. Thus, one cortical area is correlated with the performance score in at least one property. All maps were visualized using Connectome Workbench (https://www.humanconnectome.org/software/connectome-workbench).

**Table S1. Description of the cognitive tasks**

| Construct | Task | Description |
|---|---|---|
| Reasoning *(Gf)* | Raven's Progressive Matrices *(Gf1)* | Tests non-verbal reasoning using Raven's Progressive Matrices Form A. Participants are presented with spatial arrangements of patterns in a matrix, with one square missing. The missing square has to be selected out of five alternatives. In total, there are 24 items and 3 bonus items. The task stops after five incorrect answers in a row. |
| | Spatial Orientation Processing *(Gf2)* | It is measured using Variable Short Penn Line Orientation. The participants need to rotate a line to be parallel with the reference line by clicking buttons on the keyboard for clockwise or counterclockwise rotations. There are 24 items in this task. |
| | List Sorting Working Memory *(Gf3)* | The task requires participants to sort items according to life-size. Items are pictures of animals and food, accompanied by sound clips and text that name the picture. In 1-list, the participants need to sort either animals or food. In 2-list, both animals and food are displayed together, sorting should be done first for food and then for animals. |
| Comprehension Knowledge *(Gc)* | Oral Reading Recognition Test *(Gc1)* | The task tests the ability of the participants to read English and Spanish words as accurately as possible. The administrator scores them as correct or wrong. |
| | Vocabulary Comprehension *(Gc2)* | Measures general knowledge of vocabulary. Participants are to choose one out of four pictures, which most closely matches the meaning of a spoken word. |
| Memory *(Gm)* | Verbal Episodic Memory *(Gm1)* | Measures verbal episodic memory. Participants are to memorize 20 written words. Next, these words are shown together with 20 new words, needed to be distinguished. |
| | Picture Sequence Memory *(Gm2)* | Assesses visual episodic memory. A sequence of 15 pictures is presented and over two learning trials required to recall in sequence. |
| Executive Function *(EF)* | Dimensional Change Card Sort - Color *(EF1)* | Measures cognitive flexibility. Participants are presented with a series of pictures that vary along several feature dimensions. Participants who are required to match them to a target picture according to the selected feature (here it is color). After some trials, the feature is changed ("switch trial"). |

| | Dimensional Change Card Sort - Shape *(EF2)* | Same as *EF1*, but now the target feature is shape. Note that "switch" trials are also applied. |
| | Flanker Inhibitory Control and Attention Task – Congruent *(EF3)* Incongruent *(EF4* | Measures inhibition. Participants have to respond according to the direction into which an arrow is pointing. The arrow is flanked to the left and right by other arrows that point into the same direction (congruent) or the opposite direction (incongruent). |
| Mental speed *(Gs)* | Pattern Comparison Processing Speed *(Gs1)* | The task measures the speed of the participants to decide whether a pair of pictures are the same. |
| | Sustained Attention *(Gs2)* | The participants are presented with vertical and horizontal lines. The task measures the speed of the participants to press spacebar when the lines form a number or later. |
| | Relational Processing 1 *(Gs3)* | The participants are presented by pairs of objects and asked to decide whether those pairs differ in the same feature. This test is done during functional MRI scan. Gs3 relates to trials where the fMRI scan is from right to left. |
| | Relational Processing 2 *(Gs4)* | The task for *Gs4* is identical to Gs3 but relates to those trials where the fMRI scan is from left to right. |

*Note*. In the HCP project manual following task abbreviations were used: *Gf1* – PMAT24_A_CR; *Gf2* - VSPLOT_TC; *Gf3* - ListSort_AgeAdj; *Gc1* - ReadEng_AgeAdj; *Gc2* - PicVocab_AgeAdj; *Gm1* - IWRD_TOT; *Gm2* - PicSeq_AgeAdj; *Ex1* - DCCS_C_s_mean_RT_s; *Ex2* - DCCS_S_r_mean_RT_s; *Ex3* - Flanker_c_mean_RT_s; *Ex4* - Flanker_i_mean_RT_s; *Gs1* - Processing_speed_mean_RT_s; *Gs2* - Sustained_attention_median_RT_s; *Gs3* - Relational_con_mean_RT_RL_s; *Gs4* - Relational_con_mean_RT_LR_s.

**Table S2. Summary of CFA and EFA model fit estimated to identify neurometric ontological entities of human intelligence**

| Model | $\chi^2$ | CFI | RMSEA | SRMR |
|---|---|---|---|---|
| EFA of IOU Matrix from the whole sample | 35.2 | - | .000 | - |
| CFA of performance correlation matrix | 761.998 | .681 | .092 | .089 |
| CFA of IOU Matrix from the whole sample | 190.036 | .978 | .042 | .035 |

*Note*. IOU – intersection over union; CFI – Comparative Fit Index; RMSEA – Root Mean Squared Error of Approximation; SRMR – Standardized Root Mean Squared Residual

**Table S3. Standardized factor loadings and factor correlations of CFA applied in task scores correlation matrix from the whole sample**

| Tasks | Gf | Gc | Gm | EF | Gs |
|---|---|---|---|---|---|
| Raven's Progressive Matrices *(Gf1)* | .656 | | | | |
| Spatial Orientation Processing *(Gf2)* | .557 | | | | |
| List Sorting Working Memory *(Gf3)* | .473 | | | | |
| Oral Reading Recognition Test *(Gc1)* | | .822 | | | |
| Vocabulary Comprehension *(Gc2)* | | .817 | | | |
| Verbal Episodic Memory *(Gm1)* | | | .367 | | |
| Picture Sequence Memory *(Gm2)* | | | .542 | | |
| Dimensional Change Card Sort - Color *(EF1)* | | | | .800 | |
| Dimensional Change Card Sort - Shape *(EF1)* | | | | .883 | |
| Flanker Inhibitory Control and Attention Task – Congruent *(EF3)* | | | | .690 | |
| Flanker Inhibitory Control and Attention Task – Incongruent *(EF4)* | | | | .669 | |
| Pattern Comparison Processing Speed *(Gs1)* | | | | | .656 |
| Sustained Attention *(Gs2)* | | | | | .265 |
| Relational Processing 1 *(Gs3)* | | | | | .365 |
| Relational Processing 2 *(Gs4)* | | | | | .452 |
| **Factor Correlations** | *Gf* | *Gc* | *Gm* | *EF* | *Gs* |
| *Gf* | - | | | | |
| *Gc* | .790 | - | | | |
| *Gm* | .703 | .391 | - | | |
| *EF* | .302 | .253 | .348 | - | |
| *Gs* | .352 | .341 | .509 | .790 | - |

*Note:* All factor loadings and correlations are statistically significant ($p < .05$).

**Table S4. Neurometric ontologies from resampling approach**

| Realization | Sample 1 | Sample 2 |
|---|---|---|
| 1 | Factor 1: Gf1, Gf2, Gf3, Gc1, Gc2, Gm2<br>Factor 2: EF1, EF2<br>Factor 3: EF3, EF4<br>Factor 4: Gs1, Gs2, Gs3, Gs4, Gm1 | Factor 1: Gf1, Gf2, Gf3, Gc1, Gc2, Gm1, Gm2<br>Factor 2: EF1, EF2, EF3, EF4<br>Factor 3: Gs1, Gs2, Gs3, Gs4 |
| 2 | Factor 1: Gf1, Gf2, Gf3, Gc1, Gc2, Gm1<br>Factor 2: EF1, EF2, EF3, EF4, Gm2<br>Factor 3: Gs1, Gs2, Gs3, Gs4 | Factor 1: Gf1, Gf2, Gf3, Gc1, Gc2, Gs1, Gs2, Gs3, Gs4, Gm1, Gm2<br>Factor 2: EF1, EF2, EF3, EF4 |
| 3 | Factor 1: Gf1, Gf2, Gf3, Gc1, Gc2, Gm2<br>Factor 2: EF1, EF2, EF3, EF4, Gm1<br>Factor 3: Gs1, Gs2, Gs3, Gs4 | Factor 1: Gf1, Gf2, Gf3, Gc1, Gc2, Gm2<br>Factor 2: EF1, EF2, EF3, EF4<br>Factor 3: Gs1, Gs2, Gs3, Gs4, Gm1 |
| 4 | Factor 1: Gf1, Gf2, Gf3, Gc1, Gc2<br>Factor 2: EF1, EF2, EF3, EF4, Gm2<br>Factor 3: Gs1, Gs2, Gs3, Gs4, Gm1 | Factor 1: Gf1, Gf2, Gf3, Gc1, Gc2, Gm2<br>Factor 2: EF1, EF2<br>Factor 3: EF3, EF4<br>Factor 4: Gs1, Gs2, Gs3, Gs4, Gm1 |
| 5 | Factor 1: Gf1, Gf2, Gf3, Gc1, Gc2, Gm1, Gm2<br>Factor 2: EF1, EF2, EF3, EF4<br>Factor 3: Gs1, Gs2, Gs3, Gs4 | Factor 1: Gf1, Gf2, Gf3, Gc1, Gc2, Gm1<br>Factor 2: EF1, EF2, EF3, EF4, Gm2<br>Factor 3: Gs1, Gs2, Gs3, Gs4 |
| 6 | Factor 1: Gf1, Gf2, Gf3, Gc1, Gc2, Gm2<br>Factor 2: EF1, EF2, EF3, EF4, Gm1<br>Factor 3: Gs1, Gs2, Gs3, Gs4 | Factor 1: Gf1, Gf2, Gf3, Gc1, Gc2, Gm1<br>Factor 2: EF1, EF2<br>Factor 3: EF3, EF4<br>Factor 4: Gs1, Gs2, Gs3, Gs4, Gm2 |
| 7 | Factor 1: Gf1, Gf2, Gf3, Gc1, Gc2, Gm1, Gm2<br>Factor 2: EF1, EF2, EF3, EF4 | Factor 1: Gf1, Gf2, Gf3, Gc1, Gc2, Gm1, Gm2<br>Factor 2: EF1, EF2, EF3, EF4 |
| | Factor 3: Gs1, Gs2, Gs3, Gs4 | Factor 3: Gs1, Gs2, Gs3, Gs4 |
| 8 | Factor 1: Gf1, Gf2, Gf3, Gc1, Gc2, Gm1<br>Factor 2: EF1, EF2<br>Factor 3: EF3, EF4, Gm2<br>Factor 4: Gs1, Gs2, Gs3, Gs4 | Factor 1: Gf1, Gf2, Gf3, Gc1, Gc2, Gm1<br>Factor 2: EF1, EF2, EF3, EF4<br>Factor 3: Gs1, Gs2, Gs3, Gs4, Gm2 |
| 9 | Factor 1: Gf1, Gf2, Gf3, Gc1, Gc2<br>Factor 2: EF1, EF2, EF3, EF4, Gm1, Gm2<br>Factor 3: Gs1, Gs2, Gs3, Gs4 | Factor 1: Gf1, Gf2, Gf3, Gc1, Gc2, Gm2<br>Factor 2: EF1, EF2, EF3, EF4<br>Factor 3: Gs1, Gs2, Gs3, Gs4, Gm1 |
| 10 | Factor 1: Gf1, Gf2, Gf3, Gc1, Gc2, Gm1, Gm2<br>Factor 2: EF1, EF2, EF3, EF4<br>Factor 3: Gs1, Gs2, Gs3, Gs4 | Factor 1: Gf1, Gf2, Gf3, Gc1, Gc2<br>Factor 2: EF1, EF2<br>Factor 3: EF3, EF4<br>Factor 4: Gs1, Gs2, Gs3, Gs4, Gm1, Gm2 |
| 11 | Factor 1: Gf1, Gf2, Gf3, Gc1, Gc2<br>Factor 2: EF1, EF2, EF3, EF4, Gm1, Gm2<br>Factor 3: Gs1, Gs2, Gs3, Gs4 | Factor 1: Gf1, Gf2, Gf3, Gc1, Gc2, Gm1, Gm2<br>Factor 2: EF1, EF2, EF3, EF4<br>Factor 3: Gs1, Gs2, Gs3, Gs4 |
| 12 | Factor 1: Gf1, Gf2, Gf3, Gc1, Gc2<br>Factor 2: EF1, EF2, EF3, EF4<br>Factor 3: Gs1, Gs2, Gs3, Gs4, Gm1, Gm2 | Factor 1: Gf1, Gf2, Gf3, Gc1, Gc2, Gm2<br>Factor 2: EF1, EF2, EF3, EF4<br>Factor 3: Gs1, Gs2, Gs3, Gs4, Gm1 |
| 13 | Factor 1: Gf1, Gf2, Gf3, Gc1, Gc2, Gm1, Gm2<br>Factor 2: EF1, EF2, EF3, EF4<br>Factor 3: Gs1, Gs2, Gs3, Gs4 | Factor 1: Gf1, Gf2, Gf3, Gc1, Gc2, Gm2<br>Factor 2: EF1, EF2, EF3, EF4<br>Factor 3: Gs1, Gs2, Gs3, Gs4, Gm1 |
| 14 | Factor 1: Gf1, Gf2, Gf3, Gc1, Gc2, Gm1<br>Factor 2: EF1, EF2, EF3, EF4<br>Factor 3: Gs1, Gs2, Gs3, Gs4, Gm2 | Factor 1: Gf1, Gf2, Gf3, Gc1, Gc2, Gm2<br>Factor 2: EF1, EF2, EF3, EF4, Gm1<br>Factor 3: Gs1, Gs2, Gs3, Gs4 |
| 15 | Factor 1: Gf1, Gf2, Gf3, Gc1, Gc2, Gm1, Gm2<br>Factor 2: EF1, EF2, EF3, EF4<br>Factor 3: Gs1, Gs2, Gs3, Gs4 | Factor 1: Gf1, Gf2, Gf3, Gc1, Gc2<br>Factor 2: EF1, EF2, EF3, EF4, Gm1<br>Factor 3: Gs1, Gs2, Gs3, Gs4, Gm2 |
| 16 | Factor 1: Gf1, Gf2, Gf3, Gc1, Gc2<br>Factor 2: EF1, EF2, EF3, EF4, Gm1<br>Factor 3: Gs1, Gs2, Gs3, Gs4, Gm2 | Factor 1: Gf1, Gf2, Gf3, Gc1, Gc2, Gm1, Gm2<br>Factor 2: EF1, EF2, EF3, EF4<br>Factor 3: Gs1, Gs2, Gs3, Gs4 |
| 17 | Factor 1: Gf1, Gf2, Gf3, Gc1, Gc2, Gm1, Gm2<br>Factor 2: EF1, EF2, EF3, EF4<br>Factor 3: Gs1, Gs2, Gs3, Gs4 | Factor 1: Gf1, Gf2, Gf3, Gc1, Gc2, Gm1, Gm2<br>Factor 2: EF1, EF2, EF3, EF4<br>Factor 3: Gs1, Gs2, Gs3, Gs4 |
| 18 | Factor 1: Gf1, Gf2, Gf3, Gc1, Gc2, Gm2<br>Factor 2: EF1, EF2, EF3, EF4<br>Factor 3: Gs1, Gs2, Gs3, Gs4, Gm1 | Factor 1: Gf1, Gf2, Gf3, Gc1, Gc2<br>Factor 2: EF1, EF2, EF3, EF4, Gm2<br>Factor 3: Gs1, Gs2, Gs3, Gs4, Gm1 |
| 19 | Factor 1: Gf1, Gf2, Gf3, Gc1, Gc2, Gm1<br>Factor 2: EF1, EF2, EF3, EF4<br>Factor 3: Gs1, Gs2, Gs3, Gs4, Gm2 | Factor 1: Gf1, Gf2, Gf3, Gc1, Gc2, Gm1, Gm2<br>Factor 2: EF1, EF2, EF3, EF4<br>Factor 3: Gs1, Gs2, Gs3, Gs4 |
| 20 | Factor 1: Gf1, Gf2, Gf3, Gc1, Gc2<br>Factor 2: EF1, EF2, Gm1<br>Factor 3: EF3, EF4, Gm2<br>Factor 4: Gs1, Gs2, Gs3, Gs4 | Factor 1: Gf1, Gf2, Gf3, Gc1, Gc2, Gm2<br>Factor 2: EF1, EF2, EF3, EF4, Gm1<br>Factor 3: Gs1, Gs2, Gs3, Gs4 |
| 21 | Factor 1: Gf1, Gf2, Gf3, Gc1, Gc2, Gm1<br>Factor 2: EF1, EF2, EF3, EF4<br>Factor 3: Gs1, Gs2, Gs3, Gs4, Gm2 | Factor 1: Gf1, Gf2, Gf3, Gc1, Gc2, Gm2<br>Factor 2: EF1, EF2, EF3, EF4, Gm1<br>Factor 3: Gs1, Gs2, Gs3, Gs4 |
| 22 | Factor 1: Gf1, Gf2, Gf3, Gc1, Gc2, Gm2<br>Factor 2: EF1, EF2, Gm1<br>Factor 3: EF3, EF4<br>Factor 4: Gs1, Gs2, Gs3, Gs4 | Factor 1: Gf1, Gf2, Gf3, Gc1, Gc2<br>Factor 2: EF1, EF2, EF3, EF4, Gm1<br>Factor 3: Gs1, Gs2, Gs3, Gs4, Gm2 |
| 23 | Factor 1: Gf1, Gf2, Gf3, Gc1, Gc2, Gm1, Gm2<br>Factor 2: EF1, EF2<br>Factor 3: EF3, EF4<br>Factor 4: Gs1, Gs2, Gs3, Gs4 | Factor 1: Gf1, Gf2, Gf3, Gc1, Gc2, Gm1, Gm2<br>Factor 2: EF1, EF2, EF3, EF4<br>Factor 3: Gs1, Gs2, Gs3, Gs4 |
| 24 | Factor 1: Gf1, Gf2, Gf3, Gc1, Gc2 | Factor 1: Gf1, Gf2, Gf3, Gc1, Gc2, Gm1 |
| | Factor 2: EF1, EF2, EF3, EF4<br>Factor 3: Gs1, Gs2, Gs3, Gs4, Gm1, Gm2 | Factor 2: EF1, EF2, EF3, EF4, Gm2<br>Factor 3: Gs1, Gs2, Gs3, Gs4 |
| 25 | Factor 1: Gf1, Gf2, Gf3, Gc1, Gc2, Gm2<br>Factor 2: EF1, EF2, EF3, EF4<br>Factor 3: Gs1, Gs2, Gs3, Gs4, Gm1 | Factor 1: Gf1, Gf2, Gf3, Gc1, Gc2, Gm1, Gm2<br>Factor 2: EF1, EF2, EF3, EF4<br>Factor 3: Gs1, Gs2, Gs3, Gs4 |
| 26 | Factor 1: Gf1, Gf2, Gf3, Gc1, Gc2, Gm2<br>Factor 2: EF1, EF2, Gm1<br>Factor 3: EF3, EF4<br>Factor 4: Gs1, Gs2, Gs3, Gs4 | Factor 1: Gf1, Gf2, Gf3, Gc1, Gc2, Gm2<br>Factor 2: EF1, EF2, Gm1<br>Factor 3: EF3, EF4<br>Factor 4: Gs1, Gs2, Gs3, Gs4 |
| 27 | Factor 1: Gf1, Gf2, Gf3, Gc1, Gc2, Gm1, Gm2<br>Factor 2: EF1, EF2, EF3, EF4<br>Factor 3: Gs1, Gs2, Gs3, Gs4 | Factor 1: Gf1, Gf2, Gf3, Gc1, Gc2<br>Factor 2: EF1, EF2, EF3, EF4, Gm2<br>Factor 3: Gs1, Gs2, Gs3, Gs4, Gm1 |
| 28 | Factor 1: Gf1, Gf2, Gf3, Gc1, Gc2<br>Factor 2: EF1, EF2, EF3, EF4<br>Factor 3: Gs1, Gs2, Gs3, Gs4, Gm1, Gm2 | Factor 1: Gf1, Gf2, Gf3, Gc1, Gc2, Gm2<br>Factor 2: EF1, EF2, EF3, EF4<br>Factor 3: Gs1, Gs2, Gs3, Gs4, Gm1 |
| 29 | Factor 1: Gf1, Gf2, Gf3, Gc1, Gc2, Gm2<br>Factor 2: EF1, EF2, EF3, EF4<br>Factor 3: Gs1, Gs2, Gs3, Gs4, Gm1 | Factor 1: Gf1, Gf2, Gf3, Gc1, Gc2, Gm2<br>Factor 2: EF1, EF2, EF3, EF4, Gm1<br>Factor 3: Gs1, Gs2, Gs3, Gs4 |
| 30 | Factor 1: Gf1, Gf2, Gf3, Gc1, Gc2, Gs1, Gs2, Gs3, Gs4<br>Factor 2: EF1, EF2, EF3, EF4, Gm1, Gm2 | Factor 1: Gf1, Gf2, Gf3, Gc1, Gc2<br>Factor 2: EF1, EF2, Gm1<br>Factor 3: EF3, EF4, Gm2<br>Factor 4: Gs1, Gs2, Gs3, Gs4 |

Note: Blue, green and yellow shading: 2- 3-, and 4- factorial structure.

**Table S5. Standardized factor loading and factor correlations estimated by EFA applied on the IOU matrix from one of the samples**

| Tasks | Factor 1<br>(G) | Factor 2<br>(EF) | Factor 3<br>(Gs) |
|---|---|---|---|
| (Gf1) | **.61** | -.02 | -.07 |
| (Gf2) | **.67** | -.03 | -.14 |
| (Gf3) | **.28** | .04 | .04 |
| (Gc1) | **.64** | -.01 | -.10 |
| (Gc2) | **.67** | -.08 | -.02 |
| (Gm1) | **.12*** | .05 | .12 |
| (Gm2) | **.28*** | .07 | .02 |
| (EF1) | .03 | **.28** | .16 |
| (EF2) | .14 | **.26** | .19 |
| (EF3) | -.09 | **.88** | -.14 |
| (EF4) | -.01 | **.72** | -.09 |
| (Gs1) | .25 | .00 | **.27** |
| (Gs2) | .15 | -.05 | **.15** |
| (Gs3) | -.15 | -.01 | **.61** |
| (Gs4) | -.08 | -.10 | **.63** |
| **Factor Correlations** | | | |
| Factor 1 (G) | - | | |
| Factor 2 (EF) | .62 | - | |
| Factor 3 (Gs) | .67 | .51 | - |

*Note*: All factor loadings are statistically significant ($p < .05$).

The highest loadings of the task are bold faced; these loadings are shown as solid lines in Fig. 3B.

**Table S6. Standardized factor loading and factor correlations estimated by EFA applied on the IOU matrix from the whole sample**

| Tasks | Factor 1 (*G*) | Factor 2 (*EF*) | Factor 3 (*Gs*) |
|---|---|---|---|
| *(Gf1)* | **.62** | .01 | -.03 |
| *(Gf2)* | **.60** | .00 | .02 |
| *(Gf3)* | **.33** | -.01 | .12 |
| *(Gc1)* | **.73** | -.01 | -.08 |
| *(Gc2)* | **.76** | .01 | -.09 |
| *(Gm1)* | .11 | **-.03** | **.19*** |
| *(Gm2)* | **.27*** | .07 | .03 |
| *(EF1)* | .05 | **.56** | -.05 |
| *(EF2)* | .12 | **.61** | -.02 |
| *(EF3)* | -.13 | **.66** | .12 |
| *(EF4)* | .03 | **.59** | .10 |
| *(Gs1)* | .22 | .00 | **.28** |
| *(Gs2)* | .03 | .02 | **.23** |
| *(Gs3)* | -.10 | .01 | **.52** |
| *(Gs4)* | -.05 | .00 | **.54** |
| **Factor Correlations** | | | |
| **Factor 1 (*G*)** | - | | |
| **Factor 2 (*EF*)** | .62 | - | |
| **Factor 3 (*Gs*)** | .66 | .53 | - |

*Note*: All factor loadings are statistically significant ($p < .05$).

The highest loadings of the task are bold faced; these loadings are shown as solid lines in Fig. 3C.

* the task is included in the most interpretable factors. Note that these loadings on the factor are much smaller than those of the other tasks; hence, the task is not included in the factor. These loadings are shown as broken lines in Fig. 3C.